\documentclass{LMCS}
\input{amssym.def}
\input{amssym.tex}
\usepackage{enumerate}
\usepackage{hyperref}

\theoremstyle{plain}\newtheorem{lemma}[thm]{Lemma}
\newtheorem{remark}[thm]{Remark}
\newtheorem{example}[thm]{Example}
\newtheorem{observation}[thm]{Observation}
\newcommand{\R}{{\Bbb R}}
\newcommand{\N}{{\Bbb N}}

\newcommand{\Q}{{\Bbb Q}}
\newcommand{\Z}{{\Bbb Z}}
\newcommand{\Ps}{{\Bbb P}}
\newcommand{\rinf}{\rightarrow \infty}

\newcommand{\dminus}{\mbox{$\;^\cdot\!\!\!-$}}

\def\doi{5 (3:11) 2009}
\lmcsheading%
{\doi}
{1--21}
{}
{}
{Feb.~\phantom{0}9, 2009}
{Sep.~24, 2009}
{}   

\begin{document}
\title[A rich hierarchy]{A rich hierarchy of functionals of finite types}
\author{Dag Normann} \address{Department of Mathematics, The University 
of Oslo, P.O. Box 1053, Blindern N-0316 Oslo, Norway}  
\email{dnormann@math.uio.no}


\begin{abstract} We are considering typed hierarchies of total, continuous functionals using complete, separable metric spaces at the base types. We pay special attention to the so-called {\em Urysohn space}  constructed by P. Urysohn. One of the properties of the Urysohn space is that every other separable metric space can be isometrically embedded into it.

We discuss why the Urysohn space may be considered as the universal model of possibly infinitary outputs of algorithms. The main result is that all our typed hierarchies may be topologically embedded, type by type, into the corresponding hierarchy over the Urysohn space. As a preparation for this, we prove an effective density theorem that is also of independent interest.
\end{abstract}

\maketitle

\section{Introduction}\label{section2}
\subsection{Discussion}\label{subsection2.1}
\noindent One of the important paradigms of the theory of computing, and of that of computability, is that we may view algorithms and programs as data.  We are not going to challenge this paradigm. The paradigm is important practically in the design of digital computers, where everything, input data, programs and output data deep down are just sets of bits and bytes. It is also important theoretically, as it makes the existence of a universal algorithm possible and the unsolvability of the halting problem a mathematical statement.

However, using almost any programming language in practice, we have to distinguish between input data and output data, or at least declare what is what, and the programs are considered as syntactical entities that for most cases are distinguished from other kinds of data.

In this paper we will be interested in models for computing where the input data and the output data may be infinite entities. As a simple, but basic example, let us discuss  the operator
$$I(f) = \int_{0}^1f(x)d x$$
and how we should construct mathematical models for the kinds of data involved in computing integrals.  Of course, in the world of digital computers, what we will aim at is to compute the integral as a floating point value, and then the input function $f$ has to be digitally represented in some way suitable for this aim. From the point of view of numerical analysis, this is not hard to achieve, and in fact, the computability of the integral is not a big issue. However, from the point of view of a conceptual analysis it is undesirable to make the leap all at once from the set theoretical world of mathematical analysis to the finitistic world of digital computers. There are several reasons for this. We will discuss two of them:
\begin{enumerate}[(1)]
\item The step from the continuous to the discrete inevitably has to violate some of the geometrical, algebraic and analytical properties of the reals.  Unless one shows some care, it is not obvious that
$$\int_{0}^1(x^2 + x^5)dx = \int_{0}^1 x^2 dx + \int_{0}^1x^5dx$$ as numerically calculated integrals, and there are certainly going to be algebraically valid identities of this sort that are not identities in the numerical interpretation. Though the practical harm of phenomena like this may be kept at a minimum, it will be nice to have a model of computability in analysis that does not suffer from such deficiencies.
\item Though technological standards for representing various kinds of data are important for the exchange of data and programs, a conceptual analysis of computability where data of the form {\em reals} and {\em real valued functions} appear, should not be restricted to a particular standard for digitalization.
\end{enumerate}
It is of course impossible to view a real as the genuine output of an algorithm, since such outputs, even in a mathematical model, should be of a finitistic nature. An algebraic expression denoting a real may be considered to be such a finitistic entity, but then we will  be facing the problem of the meaning of calculating the value of  expressions like this. Thus, algebraic expressions are not satisfactory representations of outputs in the sense of this paper.

We will view output data as data of a particular kind, and we will advise some care in the choice of representing such data. Of course we have to consider more than just the set of data, we have to consider approximations to these data as well. But, and this is the core of our view, since it is the output data themselves that are of importance, the structure used to model the outputs of algorithms computing such data should contain the  output data we are really interested in as a kind of substructure. We may view an algorithm computing a real as running in infinite time, producing better and better approximations as time passes, but in the end, in an ideal world, and after possibly an infinite elapse of time, the output should be the real itself.

If we consider the directed complete partial ordering (dcpo) of all closed intervals ordered by reversed inclusion, we may identify a real $x$ with the closed interval $[x,x]$, and in this way, $\R$ may be viewed as a substructure of the closed interval domain.

If we want to stick to finitistic representations of approximations of reals, e.g.\  as closed intervals $[p,q]$ with rational endpoints or as closed intervals $[\frac{n}{2^k},\frac{m}{2^k}]$ with dyadic endpoints, and represent a real as an ideal of such approximations, we may canonically represent a real $x$ as the ideal of all approximating intervals with $x$ in the interior. 

This latter kind of representation is known as a {\em retract domain representation}, and we will come back to this. 

In Section 1.3 we will bring this discussion further, and draw the conclusion that the class of complete, separable metric spaces is a suitable choice of spaces modeling types of output data, or more generally, as types of {\em ground data}.

In our example of the integral, there are two other kinds of data that may concern us, those of the input function $f$ and the integration operator itself.  Here we will view functions from reals to reals as operators on ground data, and the integral as an operator at the next level, and we will use a convenient cartesian closed category containing the complete, separable metric spaces to model such classes of operators or functions.
\subsection{Outline of the paper}
We will address the following general problem;
\begin{enumerate}[$-$]
\item Given (interpretations of the expressions) $\sigma(X)$ and $\sigma(Y)$ where $X$ and $Y$ are complete, separable metric spacesand $\sigma$ is a type, how will relationships between $X$ and $Y$ give rise to relationships between $\sigma(X)$ and $\sigma(Y)$?
\end{enumerate}
In Section \ref{section4} we give a brief introduction to $qcb$-spaces and domain representations in general, and we define our "convenient" class $\mathcal Q$ of $qcb$-spaces. In Section \ref{section3} we introduce the {\em Urysohn Space} $U$ \cite{U1,U2}, and survey some of the main properties.

One of the key results of the paper is that the universality of $U$ extends to higher types. Let $\vec X = (X_1, \ldots , X_n)$ be a sequence of complete, separable metric spaces, and let $\vec U$ be the sequence of $n$ occurrences of $U$. In Section \ref{section6} we will show that if $\sigma$ is a type with $n$ free variables for base types, then we have a topological embedding of $\sigma(\vec X)$ into $\sigma(\vec U)$. If we replace occurrence no. $i$ of $U$ in $\vec U$ with a separable Banach space $Y_i$ and let each $X_i$ be homeomorphic to a closed subset of $Y_i$, our proof can also be used to prove that $\sigma(\vec X)$ can be topologically embedded into $\sigma(\vec Y)$.

An embedding-projection pair between spaces $Y$ and $X$ is normally a pair $(\varepsilon,\pi)$ of continuous functions $\varepsilon:Y \rightarrow X$ and  $\pi:X \rightarrow Y$ such that $\pi(\varepsilon(y)) = y$ for all $y \in Y$. If we have two typed structures, one with base type $Y$ and one with base type $X$, one standard way to show that we may embed the first into the second type by type is to establish an embedding-projection pair between $Y$ and $X$ and then  show that this generates an embedding-projection pair at each type.

Sometimes it is topologically impossible to have a continuous projection from $X$ to $Y$, for instance when $X = \R$ and $Y = \N$. We will see that for many important cases, we can replace the use of the projection with a sequence of probabilistic approximations. 

For spaces  in $\mathcal Q$, we introduce {\em probabilistic} embedding-projection pairs in Section \ref{section6} as a tool in the proof of the embedding theorems. 

Prior to this, we introduce the concept of density with probabilistic selection in Section \ref{section5}. In some sense, this is a warm-up for the more general concept, but it is also used as a tool for proving {\em effective density theorems} of independent interest. 

The introduction of probabilistic embedding-projection pairs, and the simpler concept  {\em density with probabilistic selection} can be seen as the main methodological contribution in the paper. The method first appeared in Normann \cite{Dag} with $\N$ for $Y$ and $\R$ for $X$.

In our setting, the proof of an effective density theorem will involve a construction of an enumeration of a topologically dense set. We will be more precise in the sequel.

The main result in Section \ref{section6} is a purely topological result, with no constructive or computable content. There is an effective, but restricted, version of the imbedding theorem from Section \ref{section6} in preparation, and the proof of the effective density theorem in Section \ref{section5} can be viewed as a preparation for this as well.

\subsection{Representing output data}\label{subsection2.2}
Blanck \cite{Jens,Jens2} carried out some pioneering work on the use of domain theory for representing topological spaces. Though we add some conceptual analysis, the technical definitions and results of this section are due to Blanck. We have to assume some familiarity with basic domain theory, see e.g.\ Abramsky and Jung \cite{AJ}, Stoltenberg-Hansen \& al.\  \cite{VIE} or Amadio and Curien \cite{AC} for introductions to the subject.
\begin{defi} \label{def1.2}
In this paper, if $X$ is a topological space, then a {\em domain representation} of $X$ will consist of a separable algebraic domain $(D,\sqsubseteq)$, a nonempty set $D^R \subseteq D$ of {\em representing objects} with the induced Scott topology and a continuous surjection $\delta:D^R \rightarrow X$.

The representation is {\em dense} if $D^R$ is a dense subset of $D$ in the Scott topology.
\end{defi}

\noindent If $(D,D^R,\delta)$ is a domain representation of $X$, and we let $D_{0}$ be the set of {\em compact} or {\em finitary} elements of $D$, we may view the elements of $D_{0}$ as {\em approximations} to the elements of $X$.

Now, if $X$ is a set of ideal {\em output data}, the elements of $D_{0}$ may be chosen as the possible intermediate approximative values obtained through the computation of some element $x$ of $X$. If we view this set of approximations as an extension of $X$, it is natural to identify each $x \in X$ with some canonical set of approximations of $x$, preferably a set that in some abstract sense can ``be computed'' from $x$ itself.  This leads us to consider the {\em retract representations},  representations where there is a continuous {\em right inverse} $\nu:X \rightarrow D^R$ of $\delta$   such that $\delta(\nu(x)) = x$ for all $x \in X$.

Finally, an output should be complete with no room for computing another output with  strictly more information.   This leads us to consider {\em upwards closed} representations, i.e.\  representations where, if $\alpha \in D^R$ and $\alpha \sqsubseteq \beta$, then $\beta \in D^R$ and $\delta(\alpha) = \delta(\beta)$.

Blanck \cite{Jens2} proved that if a topological space $X$ accepts an upwards closed retract representation, then $X$ is a regular space, and in fact it is normal. Since we restrict our attention to separable domains, $X$ will have a countable base. Then, as an application of the Urysohn metrization theorem, $X$ will be metrizable.   

We will bring this analysis a bit further.  If we use a domain representation of a space of output data, it is reasonable to assume that the set of representing objects is a closed set in the Scott topology, simply because we then work with the completion of the approximating finitary data. This leads us to consider Polish spaces, topological spaces that can be induced  from complete, separable metric spaces. In Section \ref{section3} we will introduce the Urysohn space $U$. This is universal in the sense that Polish spaces are exactly, up to homeomorphisms, the topological spaces that are closed subsets of $U$ with the induced topology.  Thus we consider $U$ to be a suitable candidate for the {\em universal datatype of output data}, or of {\em ground data} in general.

Blanck \cite{Jens} showed how we can construct a representation of each separable metric space, and this representation will indeed be an upwards closed retract representation.  Since in later sections we will want to refer to Blanck's construction, we give some of the details here.

\begin{defi}\label{definition2.1}
 Let $\langle X,d\rangle $ be a nonempty separable metric space with a countable dense subset $\{a_{n} \mid n \in \N\}$.
\begin{enumerate}[(a)] 
\item  For each $n \in \N$ and positive rational number $r$, let $$B_{n,r} = \{x \in X \mid d(x,a_{n}) \leq r\},$$ i.e.\  the {\em closed} ball of radius $r$ around $a_{n}
$.
\item Let $E_{0} $ be the set of finite sets of such closed balls, such that whenever $B_{n,p}$ and $B_{m,q}$ are in the set, then $p+q \geq d(a_{n},a_{m})$. (The balls   have at least a potential of a nonempty intersection.)
\item If $K$ and $L$ are in $E_{0}$, we let $K \sqsubseteq L$ if for
all balls $B_{n,r}$ in $K$ there is a ball $B_{m,s}$ in $L$ such that
$s + d(a_{n},a_{m}) \leq r$. 
In this case $\sqsubseteq$ will be a preorder.
(This express that $\bigcap L$ has to be a subset of $\bigcap K$, as a consequence of the triangle inequality.)
\item An ideal $\mathcal I$ in $E_{0}$ {\em represents} $x \in X$ if:
\begin{enumerate}[(i)]
\item $x \in B_{n,r}$ whenever $B_{n,r} \in K \in {\mathcal I}$.
\item For each $\epsilon > 0$ there is a $K \in \mathcal I$ such that all balls in $K$ have radii $< \epsilon$.
\end{enumerate}
\item We let $D = D^{X}$ be the ideal completion of $E_{0}$, i.e.\  the set of ideals ordered by inclusion.
Then the set of finitary elements $D_{0}$ will be the set of {\em prime ideals} in $D$.
\end{enumerate}
\end{defi}

\noindent This construction may seem unnecessarily complicated, but
something of this complexity is required if one wants to construct an
effective domain representation uniformly from an effective metric
space.

Like all domains, $D^{X}$ is equipped with the {\em Scott topology}, where a typical element of the basis will consist of all ideals containing some fixed element of  $E_{0}$. Then the map sending a representative for $x \in X$ to $x$ will be continuous.  Now, an element $x \in X$ may have more than one representative, but there will always be a least one in the inclusion ordering of the set of ideals, and in fact, the function mapping an element $x \in X$ to the least ideal representing $x$ is continuous with respect to the Scott topology.  Thus $X$ is homeomorphic to a subspace of the representing space $D^{X}$. The least ideal representing $x \in X$ will consist of all $K$ such that $x$ is in the interior of each $B_{n,r} \in K$. It is the fact that we restrict ourselves to clusters of neighborhoods where $x$ is in the interior that makes this construction continuous.

Also observe that if ${\mathcal I} \subseteq {\mathcal J}$ are two ideals, and if $\mathcal I$ represents $x \in X$, then $\mathcal J$ represents $x $. Moreover, due to the fact that metric spaces are Hausdorff, the same ideal may not represent two different elements of $X$. Blanck's construction is that of an upwards closed retract representation.

\subsubsection*{A simpler approach}
If we are not concerned with effectivity, we may construct the representing domain based on nonempty finite intersections of  closed balls. Then we automatically get a dense retract representation that is upwards closed. This approach will be taken in Section \ref{section6}.

\section{A category of $qcb$'s}\label{section4}
\noindent In this paper, we will assume that all spaces are nonempty. 

Adopting the convention from Battenfeld, Schr\"{o}der and Simpson \cite{BSS} we say that a topological space $X$ is a $qcb$-space if it is $T_{0}$ and can be viewed as the quotient space of an equivalence relation on a space with a countable base.  The corresponding category $QCB$  is, in some sense, the richest category of topological spaces that can be handled with decency using domain theory.

Schr\"oder  introduced the concept of a pseudobase, see e.g.\  \cite{Schroder2}. 
\begin{defi}
Let $X$ be a topological space. A {\em pseudobase} for $X$ is a family $\Ps$ of nonempty subsets of $X$ closed under finite nonempty intersections such that whenever $x = \lim x_{n}$ in $X$ and $x \in O$ where $O \subseteq X$ is open, there is an element $p \in \Ps$ such that
\begin{enumerate}[(i)]
\item $x \in p \subseteq O$
\item $x_{n} \in p$ for almost all $n \in \N$.
\end{enumerate}
\end{defi}
\noindent A topological space is {\em sequential} if the topology is the finest one where the convergent sequences indeed are convergent.
Schr\"oder showed that all $qcb$-spaces will admit countable pseudobases and that every $T_{0}$-space with a countable pseudobase will be a $qcb$-space.  If we consider the Blanck representation of separable metric spaces, we may form a pseudobase from the set of finitary objects, which is a set of clusters of closed balls, by letting the pseudobase elements be all nonempty intersections of such clusters.  These pseudobase elements will be closed.

In $QCB$ we use continuous functions as morphisms. Since the spaces are sequential, a function $f:X \rightarrow Y$ is continuous if and only if it maps a convergent sequence and its limit point to a convergent sequence and its limit point.

We are going to work within a  subcategory $\mathcal Q$ of  $QCB$:
\begin{defi} Let $\mathcal Q$ be the class of sequential Hausdorff spaces that permit a countable pseudobase of closed sets.\end{defi}
By the observation above, every complete, separable metric space will be in $\mathcal Q$. We will show that $\mathcal Q$ is closed under the  function space operator used in $QCB$, and $\mathcal Q$ will then be a convenient subclass of $qcb$ for us to work  with.
\begin{remark}{\em
In  \cite{Math}, Schr\"{o}der works with a similar category, requiring that there is a pseudobase of functionally closed sets (see Definition \ref{4.8}), but not insisting on the spaces being Hausdorff. It is open whether the subcategory of Hausdorff spaces in Schr\"{o}der's category coincides with ${\mathcal Q}$.}\end{remark}
For our next result, we need the concept of an admissible domain representation due to Hamrin \cite{Hamrin}, based on a similar concept due to Schr\"oder \cite{Schroder, Schroder2}, see also Weihrauch \cite{Weihrauch.2000}:
\begin{defi} Let $\langle D , D^R , \delta \rangle$ be a representation of the space $X$, see Definition \ref{def1.2}. We call the representation {\em admissible} if for every dense representation $\langle E,E^R,\pi \rangle$ of a space $Y$  and every continuous function $f:Y \rightarrow X$ there is a continuous function $\phi:E \rightarrow D$ such that $\phi$ maps $E^R$ into $D^R$ and such that $$\delta(\phi(e)) = f(\pi(e))$$ for all $e \in E^R$.\end{defi}
\begin{remark}\label{remark7} {\em If $\langle D, D^R, \delta\rangle$ is an admissible representation of $X$ and $x = \lim_{n \rinf}x_{n}$, there will be a convergent sequence $\alpha = \lim_{n \rinf}\alpha_{n}$ in $D^R$ with $x = \delta (\alpha)$ and $x_{n} = \delta(\alpha_{n})$ for each $n \in \N$.

We call this a {\em lifting} of the convergent sequence, and the existence of a lifting is easy to prove given an admissible representation.  This is a standard observation.}\end{remark} 
\begin{lemma} \label{lemma1} Every space in $\mathcal Q$ has an upwards closed admissible representation. \end{lemma}
\proof
Let $X \in {\mathcal Q}$ and let $\Ps$ be a countable pseudobase of closed subsets of $X$. We apply the argument from Hamrin \cite{Hamrin}, and assume w.l.o.g.\  that $\Ps$ is closed under finite unions. Then the ideal completion $\langle D,\sqsubseteq\rangle$ of $\langle \Ps, \supseteq\rangle$ offers an admissible representation of $X$, where each $x \in X$ is represented by the elements of $$D^R_{x} = \{\alpha \in D \mid \forall p \in \alpha (x \in p) \wedge \forall O \;{\rm open} (x \in O \Rightarrow \exists p \in \alpha ( p \subseteq O))\}.$$
By Hamrin \cite{Hamrin} this is an admissible representation, and we are left with showing that $D^R_{x}$ is upwards closed.

If $\alpha \in D^R_{x}$ and $\alpha \subseteq \beta \in D$, the second requirement for $\beta \in D^R_{x}$ is trivially satisfied. Now, let $q \in \beta$ and assume that $x \not \in q$. Then $x \in X \setminus q$, which is open, so $$\exists p \in \alpha (x \in p \subseteq X \setminus q).$$
Then $p \cap q \in \beta$ since $\beta  $ is an ideal. But $p \cap q = \emptyset$ and $\beta$ will only contain nonempty sets. This is a contradiction, so $x \in q$.\qed
These spaces are sequential,  which means that the topology will be the finest topology where all convergent sequences do converge.  This offers a natural  topology on the function spaces $X \rightarrow Y$ of continuous functions, induced by the limit-space construction
$$f = \lim_{n \rightarrow \infty}f_{n }\Leftrightarrow \forall( x = \lim_{n \rightarrow \infty} x_{n})
(f(x) = \lim_{n \rightarrow \infty}f_{n}(x_{n})).$$
\begin{lemma}\label{lemma22} If $X$ and $Y$ are in $\mathcal Q$, then $X \rightarrow Y \in {\mathcal Q}$.
\end{lemma}
\proof
Let $p_{1} , \ldots , p_{n}$ be closed pseudobase elements in $X$ and $q_{1}, \ldots , q_{n}$ be closed pseudobase elements in $Y$ such that for all $K \subseteq \{1 , \ldots , n\}$ , $$\bigcap_{k \in K}p_{k} \neq \emptyset \Rightarrow \bigcap_{k \in K}q_{k} \neq \emptyset.$$
Let $$P_{\{\langle p_{1},q_{1}\rangle , \ldots \langle p_{n},q_{n}\rangle\}} = \{f \mid \forall k \leq n (f[p_{k}] \subseteq q_{k})\}.$$
The nonempty such sets  will form a pseudobase of closed sets for $X \rightarrow Y$. $X \rightarrow Y$ is clearly Hausdorff. \qed
\begin{remark}{\em  We do not use that $X$ is in $\mathcal Q$, only that $X$ is a $qcb$.}\end{remark}
Still using continuous functions as morphisms, we may view $\mathcal Q$ as a category.  Our key examples will be the spaces we may obtain from complete, separable nonempty metric spaces closing under the function space construction.  It is known, see Schr\"oder \cite{Sch2}, that these spaces need not be regular (or normal) spaces.  We will be interested in the finest regular (or normal, this amounts to the same in this case) subtopology of the sequential one:

\begin{defi}\label{4.8} Let $X \in {\mathcal Q}$ and let $A \subseteq X$.

We say that $A$ is {\em  functionally closed} if there is a continuous map $f:X \rightarrow [0,1]$ such that $$x \in A \Leftrightarrow f(x) = 0.$$
The complement of a functionally closed set is {\em functionally open}.
\end{defi} 
\begin{remark}{\em 
This is standard terminology from general topology. Functionally closed sets are also known as {\em zero-sets}.}\end{remark}
It is not hard to show that the functionally open sets form a regular subtopology on $X$. The fact that the topology on $X$ is hereditarily Lindel\"of, i.e.\ that every open covering of a subset accepts a countable subcovering, is useful in showing that this class is closed under arbitrary unions.
These concepts will be important in Section \ref{section6}.

In the sequel we will use the fact that if $X \in {\mathcal Q}$ and $\Ps$ is a pseudobase for $X$ consisting of closed sets, and $Y \subseteq X$, then
$$\{p\cap Y \mid p \in \Ps \wedge p \cap Y \neq \emptyset\}$$ forms a pseudobase of closed sets for $Y$.

In this paper, we let $V_{1}, \ldots , V_{k}$ be {\em formal variables} for complete, separable metric spaces, and we define the formal {\em types} as the least set of expressions containing each variable $V_{i}$ and closed under the syntactical operation $\sigma,\tau \vdash (\sigma \rightarrow \tau)$.

If $X_{1} , \ldots, X_{k}$ are separable, complete metric spaces and $\sigma$ is a type in the variables $V_1, \ldots , V_k$, its interpretation $\sigma(X_{1}, \ldots ,X_{k})$ is given in $\mathcal Q$.

It is easy to see that if each $X_{i}$ is nonempty, then $\sigma(X_{1}, \ldots , X_{k})$ is nonempty.

\section{The Urysohn Space}\label{section3}
\noindent In Section \ref{section2} we were primarily interested in mathematical models for data-types where the data could be viewed as the ultimate outputs of algorithms running in infinite time, and we observed that we may use Polish spaces or separable, complete metric spaces for this purpose. Given some metric spaces as basic data-types, we will then be interested in derived data-types, where the objects in a sense are operators with ultimate values in metric spaces.  In this paper, we will be mainly interested in hereditarily total objects of this kind, but of course, if one is interested in functional programming where such base types are involved, the hereditarily partial operators are essential for the construction of denotational semantics. 

Urysohn \cite{U1,U2} showed that there is a richest separable metric space, the so-called {\em Urysohn space}, and the main aim of this paper is to show that any  space of hereditarily total continuous functionals over any set of complete separable metric spaces can be topologically embedded into a space of functionals of the same type, but now over just the Urysohn space.

In order to be able to prove our results, we have to refer to the basic original properties of this space and to some of the more recent results about it.
\begin{defi} Let $X$ be a metric space. We call $X$ {\em finitely saturated} if whenever $K \subseteq L$ are finite metric spaces, and $\phi:K \rightarrow X$ is a metric-preserving map, then $\phi$ can be extended to a metric-preserving map $\psi$ from $L$ to $X$.\end{defi}
\begin{remark}{\em The word {\em saturated} is common in model theory for this kind of phenomenon, so we adopt it here.}\end{remark}
Urysohn proved that there exists a complete, separable metric space $U$ that is finitely saturated, and that, up to isometric equivalence, there is exactly one such space. This space is known as the {\em Urysohn space}.

Urysohn gave an explicit construction of $U$, as the completion of a countable metric space where all distances are rational numbers, and which is saturated with respect to pairs of finite spaces with rational distances. He showed that if $X$ is a metric space, $x_{1}, \ldots , x_{n}$ are elements of $X$ and $\{x_{1}, \ldots , x_{n}\}$ is extended to a metric space $\{x_{1}, \ldots , x_{n},y\}$ where $y$ is a new element with distance $d(x_i,y)$ to each $x_i$, we may consistently define a distance from $y$ to any element $x \in X$ by
$$d(x,y) = \min\{d(x,x_{i})+d(x_{i} , y)\mid 1 \leq i \leq n\}.$$ By iterating this construction using some book-keeping that ensured that all rational one point extensions of finite subspaces of the set under construction will be taken care of, he constructed the dense subset $U_0$ of $U$.

There are both effective (Kamo \cite{kano}) and constructive (Le\v{s}nik \cite{student1,student2}) versions of the main results of Urysohn.  Since effectivity is essential for our results in Section \ref{section5}, we will give a brief introduction to what we mean by effectivity.

\begin{defi}
A real $x$ is {\em computable} if there is a {\em fast converging} computable sequence $\{x_i\}_{i \in \N}$ of rationals with $x$ as the limit, where {\em fast converging} means that $|x_n - x| \leq 2^{-n}$ for all $n$.

A sequence $\{x_n\}_{n \in \N}$ of reals is computable if there is a computable map $\gamma$ of $\N$ into the set of fast converging sequences of rational numbers such that $x_n = \lim_{ n \rinf} \gamma(n)$ for each $n$.

A metric space $(X,d)$ is {\em effective} if there is an enumeration $\{r_i\}_{i \in \N}$ of a dense subset of $X$ such that the map
$$(i,j) \mapsto d(r_i,r_j)$$ is computable.

If $(X,d,\{r_i\}_{i \in \N})$ and $(Y, d' , \{s_j\}_{j \in \N})$ are two effective metric spaces, then an {\em effective embedding} of $X$ into $Y$ is a computable map $$i \mapsto \{j_{n,i}\}_{n \in \N}$$ such that
\begin{enumerate}[(i)]
\item $\{s_{j_{n,i}}\}_{n \in \N}$ is fast converging to some $y_i \in Y$ for each $i \in \N$.
\item $d(r_i,r_j) = d'(y_i,y_j)$ for all $i$ and $j$ in $\N$.
\end{enumerate}
\end{defi}

\noindent A careful reading of Urysohn's construction tells us that $U$ is effective in this sense.

In order to prove that the completion $U$  of $U_0$ is saturated, we will start with elements $u_1 , \ldots , u_k$ in $U$ and  {\em requirements} $d(u_i,x) = a_i$ consistent with the axioms of metric spaces for $i = 1 , \ldots , k$, and we have to prove that there is some $u \in U$ satisfying these requirements.

The proof can be made effective in the following sense:

If we represent $u_1, \ldots , u_k$ with fast converging sequences from $U_0$ and $a_1 , \ldots , a_k$ with fast converging sequences from $\Q$, we can construct a fast converging sequence from $U_0$ converging to a desired $u$. There are details to be filled in here, of course.

Then, by an application of the recursion theorem, we see that every effective metric space $(X,d)$ can be effectively embedded into $U$. Thus we have:

\begin{thm}\label{Kano} Every separable metric space $X$ can be isometrically embedded into the Urysohn space, and if $X$ is an effective space, the embedding can be made effective.\end{thm}
We of course have that the image of $X$ will be functionally closed (i.e.\  just closed) in $U$ exactly when $X$ is complete, and this is the reason for why we restrict our attention to complete, separable metric spaces in the technical sections of the paper.

There has been a renewed interest in the Urysohn space over the last 25 years.  One result in particular is of importance to us:

Uspenskij \cite{Usp} shows that $U$ as a topological space is homeomorphic to the Hilbert space $l_{2}$, and thus to any separable Hilbert space of infinite dimension. Uspenskij depends on a characterization of the class of topological spaces homeomorphic to Hilbert spaces due to Toru\'{n}czyk \cite{Tor}.

The combined  Toru\'{n}czyk - Uspenskij proof gives us no information about whether this result is constructive in any sense.

In the case of choosing a domain representation for the Urysohn space, the two approaches discussed in Section \ref{section2} are equivalent. This can be seen from the following
\begin{observation}
Let $U$ be the Urysohn space and let $B_1, \ldots , B_n$ be a family of closed balls where $B_i$ has radius $r_{i}$ and center in $a_{i}$, and assume that none of the balls are contained in the interior of any of the others.

Then the following are equivalent:
\begin{enumerate}[\em(1)]
\item $B_1 \cap \cdots \cap B_{n} \neq \emptyset.$
\item $r_i - r_j \leq d(a_{i},a_{j}) \leq r_{i}+r_{j}$ for all $i,j$ with $1 \leq i,j \leq n$.
\end{enumerate}
\end{observation}
\noindent$(2) \Rightarrow (1)$ is a consequence of saturation, there is an element in the intersection of the spheres of radius $r_{i}$ around $a_{i}$ for $i = 1 , \ldots , n$.\medskip

\noindent$(1) \Rightarrow (2)$ is a consequence of the triangle inequality.

\section{Effective density theorems}\label{section5} 
\noindent The underlying problem in this section is when we may effectively enumerate a dense subset of the set of continuous functionals of a fixed type using effective, separable metric spaces at base types. 
We will not answer this problem completely, but that the answer is not  ``{\em always}" is demonstrated by the following example, where we construct an effective metric space $A$ such that there is no effective enumeration of a dense subset of $A \rightarrow \N$:

\begin{example}{\label{example}\em Let $A \subseteq \N$ be recursively enumerable but not computable, and let $f:\N \rightarrow \N$ be a computable 1-1 enumeration of $A$.

We will construct an effective subspace of the Banach space $l_{\infty}$ of all bounded sequences of reals.

Let $a<b$ be reals, and let $[a,b]_{n}$ be those $g\in l_{\infty}$ where $g(n) \in [a,b]$ and $g(m) = 0$ for $m \neq n$.

Let $X$ consist of the constant $0$ together with all $[0,3]_{n}$ for $n \in A$ and all $[1,3]_{n}$ for $n \not \in A$.

It is easy to see that we can effectively enumerate a dense subset of $X$ with a computable metric, using a stage $m$ where $f(m) = n$ to decide to extend the ongoing sub-enumeration of $[1,3]_{n}$ to a sub-enumeration of $[0,3]_{n}$. Thus $X$ is an effective metric space.

If we have an effectively enumerated dense set $\{g_{n}\mid n \in \N\}$ of total functions in $X \rightarrow \N$, we see from the obvious connectedness-properties of $X$ that
$$n \not \in A \Leftrightarrow \exists m ( g_{m}(\lambda k.0) \neq g_{m}(n \mapsto 2))$$ where $n \mapsto 2$ is the element in $[1,3]_{n}$ that takes the value 2 on $n$.

This would imply that $A$ is computable, so there is no such sequence $\{g_{n}\}_{n \in \N}$.
}\end{example}
\noindent As a tool of independent interest, we develop the concept of {\em density with probabilistic selection}. Probabilistic selection from a dense set may replace the use of a continuous or even effective selection of a sequence from a dense set converging to a given point, when such selections are topologically impossible.

Let $A = \{a_{1} , \ldots , a_{n}\}$ be a finite set.  A {\em probability distribution} on $A$ is a map $m :A \rightarrow \R_{[0,1]}$ such that $$\sum_{k \leq n} m(a_{k}) = 1.$$
A probability distribution on a finite set $A$ induces a probability measure on the powerset of $A$, and we will not distinguish between the distribution and the induced measure.

We let $PD(A)$ be the set of probability distributions on $A$, where we assume that $A$ comes with an enumeration. $PD(A)$ can be viewed as a convex subspace of a finite dimensional Euclidean space, and thus $PD(A)$ has a canonical topology. $PD(A)$ can actually be identified with the standard simplex in $\R^n$.
\begin{defi} Let $ \{\langle A_{n} ,\nu_{n}, m_{n}\rangle\}_{n \in \N}$ be a sequence of finite sets $A_{n}$, maps $\nu_{n}:A_{n} \rightarrow X$ into a space $X \in {\mathcal Q}$  together with probability distributions $m_{n}$ on each $A_{n}$.

Let $x \in X$.
We say that $$x = \lim_{n \rinf}\nu_{n}[A_{n}]\; mod\; m_{n}$$ if whenever we for each $n \in \N$ select an $a_{n} \in A_{n}$ with $m_{n}(a_{n}) > 0$ , then $x = \lim_{n \rinf} \nu_{n}(a_{n})$.\end{defi}
We write $\nu_{n}[A_{n}]$ since it is actually the {\em image} of $A_{n}$ under $\nu_{n}$ that converges modulo the sequence of measures.
\begin{defi}\label{def4.3} Let $X$ be in $\mathcal Q$.
$X$ satisfies {\em density with probabilistic selection} if there are
\begin{enumerate}[(i)]
\item a sequence $\{A_{n}\}_{n \in \N}$ of finite sets together with maps $\nu_{n}:A_{n} \rightarrow X$
\item a sequence of continuous maps
$$ \mu_{n}:X \rightarrow PD(A_{n})$$
\end{enumerate}
such that  for each $x \in X$:
$$x = \lim_{n \rinf}\nu_{n}[A_{n}] \;mod\; \mu_{n}(x).$$
When this is the case, we call $\{\langle A_{n},\nu_{n}, \mu_{n}\rangle\}_{n \in \N}$ {\em a probabilistic selection on $X$}.
 \end{defi}
If  $\{\langle A_{n},\nu_{n}, \mu_{n}\rangle\}_{n \in \N}$ is a probabilistic selection on $X$, then $\bigcup_{i \in \N}\nu_{n}[A_{n}]$ will be dense in $X$ and for every $x \in X$, the set of sequences $$\{a_n\}_{n \in \N} \in \prod_{n \in \N}A_n$$ such that $x = \lim_{n \rinf}\nu_n(a_n)$ will have measure 1 in the product measure $\prod_{n \in \N}\mu_n(x)$.
\begin{remark}{\em
This concept will be an important tool in showing density theorems.  In order to prove embedding theorems, we will extend this concept in Section \ref{section6} to what we will call {\em a probabilistic projection}.

 In our applications, $X$ will be a space $$X = 
 \sigma(X_1, \ldots,X_k)$$ where each $X_i$ is a complete, separable metric space.  Then $A_{n}$ will consist of finite functionals of the same type, where the base types are interpreted as finite subsets of the metric spaces in question.  Then $\nu_{n}$ represents a way to embed these finitary functionals into the space of continuous functionals. } \end{remark}

\begin{lemma} \label{lemma5.2}Let $X$ be a separable metric space. Then $X$ satisfies density with probabilistic selection.\end{lemma}
\proof
Let $d$ be the metric on $X$, and let $\{a_{0} , a_{1}, \ldots\}$ be a countable dense subset of $X$. Let $A_{n} = \{a_{0} , \ldots , a_{n}\}$ and let  $\nu_{n}$ be the inclusion function from $A_{n}$ to $X$.
\begin{enumerate}[$-$]
\item For any $x \in X$, let $d(A_{n} , x) = \min\{d(x , a_{i})\mid i \leq n\}$.

\item If $u$ and $v$ are non-negative reals, let $u \dminus v = \max\{u-v , 0\}$.

\item For each $x \in X$ and $a \in A_{n}$, let
$$\mu_{n}(x)(a) = \frac{(d(x,A_{n}) + \delta_{n}) \dminus d(x,a)}{\sum_{b \in A_{n}}[(d(x,A_{n}) + \delta_{n}) \dminus d(x,b)]},$$
where $\delta_{n}$ is the minimum of $2^{-n}$ and all distances $d(a,b)$ for $a \neq b$ in $A_{n}$.
\end{enumerate}
The required properties are easy to verify.\qed
\begin{defi} Let $X \in {\mathcal Q}$.
We say that $X$ is {\em semiconvex} if for every finite set $$A = \{a_{1} , \ldots , a_{n}\} $$ and map $\nu:A \rightarrow X$, there is a continuous $$h_{A,\nu}:PD(A) \rightarrow X$$ such that the following holds:
Whenever
\begin{enumerate}[$-$]
\item $A_{n} $ is finite for each $n \in \N$,
\item $\nu_{n}:A_{n} \rightarrow X$ for each $n \in \N$,
\item $m_{n} \in PD(A_{n})$ for each $n \in \N$,
\item $x \in X$ is such that $$x = \lim_{n \rinf} \nu_{n}[A_{n}]\;mod\;m_{n}$$
for each $n \in \N$,
\end{enumerate}
then$$x = \lim_{n \rinf}h_{A_{n},\nu_{n}}(m_{n}).$$
\end{defi}
\begin{lemma} \label{lemma5.5new}The Urysohn space $U$ is semiconvex.\end{lemma}
\proof
 Let $A =  \{a_{1}, \ldots , a_{n}\}$ be finite and let $\nu:A \rightarrow U$.
 Let $v_{i} = \nu(a_{i})$ and let $V = \{v_{1} , \ldots , v_{n}\}$.
We may let $\phi$ embed $V$ isometrically into $\R^n\;$ with the max-norm and we may let $\psi$ embed $\R^n$ with the max-norm isometrically into $U$  such that $\psi(\phi(v_{i})) = v_{i}$ for all $i \leq n$.
Then let $$h_{A,\nu}(m) = \psi(\sum_{i = 1}^n m(a_{i})\cdot \phi(v_{i})),$$
where the algebra takes place in $\R^n$.
It is easy to see that this works.\qed

\begin{remark}\label{remark5.5} {\em Clearly, every Banach space $X$ is semiconvex. If
 $A = \{a_1 , \ldots , a_n\}$ and $\nu:A \rightarrow X$ we let
 $$h_{A,\nu}(m) = \sum_{i = 1}^n m(a_i)\cdot \nu(a_i).$$}\end{remark}

\begin{thm} \label{theorem5.6}Let $X$ and $Y$  be  $\mathcal Q$-spaces that satisfy density with probabilistic selection, and assume that $Y$ is semiconvex.
Then $X \rightarrow Y$ satisfies density  with probabilistic selection.
\end{thm}
\proof
Let  $\{A_{n}\}_{n \in \N}$ be a sequence of finite sets with maps  $\nu_{n}:A_{n}\rightarrow X$ and continuous functions $$\mu_{n}:X \rightarrow PD(A_{n})$$ forming a probabilistic selection.

Let $C_{n}$ with $\theta_{n}:C_{n} \rightarrow Y$ and $ \lambda_{n}:Y \rightarrow PD(C_{n})$ for each $n \in \N$ witness that $Y$ satisfies density with probabilistic selection. Let $h_{n}:PD(C_{n}) \rightarrow Y$ be derived from the map $C \mapsto h_{C}$ witnessing that $Y$ is semiconvex.

Let $B_{n} = A_{n} \rightarrow C_{n}$ and let $\phi \in B_{n}$. First we will see how to construct a continuous $\nu_{n}^*(\phi):X \rightarrow Y$:

Let $x \in X$. For each $c \in C_{n}$ let $ \mu^{-1}_{n,x,\phi}(c)$ be defined as
$$\mu^{-1}_{n,x,\phi}(c) =  \mu_{n}(x)(\phi^{-1}(\{c\}))$$ and let $$\nu^*_{n}(\phi)(x) = h_{n}(\mu^{-1}_{n,x,\phi}).$$
We will see how the sets $B_{n}$ together with the maps $\nu^*_{n}$ from $B_{n}$ to $X \rightarrow Y$ can be organized to a probabilistic selection.

Let  $f:X \rightarrow Y$ be continuous. We will define the probability distribution $ \eta_{n}(f)$ on $B_{n}$ as a product measure and prove the required properties. Let
$$ \eta_{n}(f)(\phi) = \prod_{a \in A_{n}} \lambda_{n}(f(a))(\phi(a)).$$
$\eta_{n}(f)$ will be a probability distribution since it is the finite full product of probability distributions.
We have to show\medskip

\noindent{\em Claim:}\ 
Let $f = \lim_{n \rinf}f_{n}$ in $X \rightarrow Y$ and
assume that $$\eta_{n}(f_{n})(\phi_{n}) > 0$$ for each $n$.
Then $f = \lim_{n \rinf}\nu^*_{n}(\phi_{n}).$\medskip

\noindent{\em Proof of Claim:}\
Since we are operating in the category of sequential topological spaces, this amounts to showing that if $x = \lim_{n \rinf} x_{n}$ in $X$, then $f(x) = \lim_{n \rinf} \nu^*_{n}(\phi_{n})(x_{n})$ in $Y$.

This will follow from the construction of the $\nu^*_{n}$'s, the properties of the $h_{n}$'s and the following\medskip

\noindent{\em Subclaim:}
$f(x) = \lim_{n \rinf} \theta_{n}[C_{n}]\;mod\;\mu^{-1}_{n,x_{n},\phi_{n}}$.\medskip

\noindent{\em Proof of Subclaim:}
Let $\mu^{-1}_{n,x_{n},\phi_{n}}(c_{n}) > 0$ for each $n$.
Then there is an $a_{n} \in A_{n}$ with $\phi_{n}(a_{n}) = c_{n}$ and $ \mu_{n}(x_{n})(a_{n}) > 0$.

$x = \lim_{n \rinf}\nu_n(a_{n})$ since we have probabilistic selection on $X$, so
$$f(x) = \lim_{n \rinf}f_{n}(\nu_{n}(a_{n})).$$
Since $ \eta_n(f_{n})({\phi_{n}}) > 0$ we must have that $$ \lambda_{n}(f_{n}(a_{n}))(\phi_{n}(a_{n})) > 0$$ so $$f(x) = \lim_{n \rinf}\theta(\phi_{n}(a_{n})),$$ or, in other words
$$f(x) = \lim_{n \rinf} \theta_{n}(c_{n}).$$ This ends the proof of the subclaim, the claim and the theorem.
\qed
The proof of Theorem \ref{theorem5.6} is effective in the sense that we have given explicit constructions of all items involved. In particular this means that if we start with effective domain representations where the extra parameters ($\nu$, $\mu$ etc.) are effective, then $X \rightarrow Y$ will be represented over an effective domain, with effective density with probabilistic selection.

 We have not proved that $X \rightarrow Y$ will be semiconvex under the assumptions of Theorem \ref{theorem5.6}. In order to make use of Theorem \ref{theorem5.6}  as an induction step, we in addition need the following observation:
\begin{observation} \label{obs1}If $X$ and $Y$ are in $\mathcal Q$ and satisfy density with probabilistic selection, then so does $X \times Y$, where $X \times Y$ is the sequentialisation of the product topology on the set $X \times Y$ (i.e.\  the product in  $QCB$).\end{observation}
Clearly this observation extends to finite cartesian products.
Using standard currying of types, Observation \ref{obs1} and Theorem \ref{theorem5.6} for the induction step, we then get

\begin{thm}\label{corollary5.7} Let each of $X_{1} , \ldots, X_{k}$  be either an effective Banach space or   the Urysohn space $U$, let $\sigma$ be a type and let  $X= \sigma(X_1, \ldots,X_k)$. Then there is an effective sequence of finite sets $A_n$, an effective sequence of finite maps $\nu_n:A_n \rightarrow X$ and an effective sequence of continuous maps $\mu_{n}:X \rightarrow PD(A_{n})$ such that $\{\langle A_{n},\nu_{n},\mu_{n}\rangle\}_{n \in \N}$ is a probabilistic selection on $X$.\end{thm}
Our starting point was the search for an effective enumeration of a dense subset of some spaces of functionals of a given type.  We have obtained
\begin{cor}
 Let each of $X_{1} , \ldots, X_{k}$  be either an effective  Banach space or the Urysohn space $U$. Let $\sigma$ be a type and let $X = \sigma(X_1 , \ldots , X_k)$.

 Then there is an effective enumeration of a dense subset of $X$.

\end{cor}
\proof
Recall the comment after Definition \ref{def4.3} and then use Theorem \ref{corollary5.7}.\qed

\section{An embedding theorem}\label{section6}
\noindent In this section we will prove a theorem that is strictly topological in formulation, but where the motivation for proving it comes from the wish to understand the nature of the spaces used in the semantics of functional programming.

We will prove the following:
\begin{thm}\label{theorem6.1} Let $\sigma$ be a type in the variables $V_{1}, \ldots , V_{k}$ and let $X_{1} , \ldots , X_{k}$ be complete, separable metric spaces. 

Then $\sigma(X_{1} , \ldots , X_{k})$ is homeomorphic to a functionally closed set in $\sigma(U, \ldots , U)$,  where $U$ is the Urysohn space.\end{thm}

In order to prove this theorem, we have to work with a combination of the concept of an embedding-projection pair and probabilistic selection as defined in Section \ref{section5}. 
\begin{defi} 
A {\em probabilistic embedding-projection pair} between $Y$ and $X$ consists of
\begin{enumerate}[(i)]
\item A sequence $\{A_{n}\}_{n \in \N}$ of finite sets together with maps $\nu_{n}:A_{n} \rightarrow Y$.
\item A continuous map $\varepsilon:Y \rightarrow X$ onto a functionally closed subset of $X$.
\item A sequence of continuous maps
$$\mu_{n}:X \rightarrow PD(A_{n})$$
\end{enumerate}
such that: \begin{enumerate}[$-$]
\item When $x = \lim_{n \rinf}x_{n}$ in $X$ with $x = \varepsilon(y)$ for some $y \in Y$, and $a_{n} \in A_{n}$  for each $n \in \N$ is such that $\mu_{n}(x_{n})(a_{n}) > 0$, we have that $y = \lim_{n \rinf}\nu_{n}(a_{n})$.
\end{enumerate}
We will call a sequence $\{\langle A_{n},\nu_{n},\mu_{n}\rangle\}_{n \in \N}$ like this a {\em probabilistic projection}.\end{defi}
In a probabilistic  embedding-projection  pair as above, we clearly have that $\varepsilon$ is injective.
\begin{lemma}\label{lemma6.2new}Let $X$ and $Y$  be  complete separable metric spaces, and let $Y$ be isometric to a subspace of $X$ via $\varepsilon:Y \rightarrow X$. 
Then $\varepsilon$ is the embedding-part of a probabilistic embedding-projection pair between $Y$ and $X$.
\end{lemma}
\proof
We use the construction from the proof of Lemma \ref{lemma5.2}, replacing the enumeration of a dense subset of $X$ with an enumeration of a dense subset of $Y$, and relating $x \in X$ to the $\varepsilon$-range of finite parts of the dense subset of $Y$. There are no new technical aspects of the proof. Note that since $Y$ is complete, the image of $\varepsilon$ is closed in $X$,  and thus functionally closed.
\qed
The key lemma in proving Theorem \ref{theorem6.1} is 
\begin{lemma}\label{lemma6.2} Let $X\in {\mathcal Q}$, $Y$ homeomorphic to a functionally closed set in $X$ via an embedding $\varepsilon:Y \rightarrow X$.
Let $A \subseteq U$ be a closed subset of the Urysohn space $U$.

If $\varepsilon$ is the embedding-part of a probabilistic embedding-projection pair between $Y$ and $X$,  then $Y \rightarrow A$ is homeomorphic to a functionally closed set  $Z$ in $X \rightarrow U$ admitting a probabilistic embedding-projection pair  between $Y \rightarrow A$ and  $X \rightarrow U$.\end{lemma}
\begin{remark}{\em  We restrict ourselves to $\mathcal Q$ everywhere, also in cases where the proof works for $qcb$-spaces in general, or even in a greater generality.}\end{remark}
Theorem \ref{theorem6.1} is proved by induction on the type, using
Lemma \ref{lemma6.2new} in the base case and Lemma \ref{lemma6.2} in
the induction step. For the induction step, we will also need Lemma
\ref{lemma6.3} handling cartesian products.

\proof [Proof of Lemma \ref{lemma6.2}]
For each $n$ let $A_{n} \subseteq Y$ be finite, $\nu_{n}:A_{n} \rightarrow Y$  and $\mu_{n}:X \rightarrow PD(A_{n})$ be continuous such that the sequences form a probabilistic projection.

Let $f:X \rightarrow [0,1]$ be continuous such that $$\varepsilon[Y] = f^{-1}(\{0\}).$$

First we will show how to embed $Y \rightarrow A$ into $X \rightarrow U$. We will use that $U$ is homeomorphic to $l_{2}$, see Uspenskij \cite{Usp}, and the linear operations below are carried out via this homeomorphism.

Let $g:Y \rightarrow A$ be continuous and let $x \in X$. Let
\[\varepsilon^*(g)(x) =
  \begin{cases}
    g(\varepsilon^{-1}(x))&
    \hbox{if $x \in \varepsilon[Y]$}\cr
    {\displaystyle(1-\lambda) \sum_{a \in A_{n}}\mu_{n}(x)(a)\cdot g(\nu_{n}(a)) +
    \lambda \sum_{b \in A_{n+1}}\mu_{n+1}(x)(b)\cdot g(\nu_{n}(b))}\cr
    \quad\hbox{where $n \in \N$ and $\lambda \in [0,1)$ are unique such
          that $f(x) = \frac{1}{n+\lambda},$}&
    \hbox{otherwise}\cr
  \end{cases}
\]
We have to  show that $\varepsilon^*(g)\in X \rightarrow U$ is continuous and that $$ \varepsilon^* \in (Y \rightarrow A)\rightarrow (X \rightarrow U)$$ is continuous.

Since we are working with sequential spaces, this amounts to showing\medskip

\noindent{\em Claim 1:}
If $g = \lim_{n \rinf} g_{n}$ in $Y \rightarrow A$ and $x = \lim_{n \rinf} x_{n}$ in $X$  then
$$\varepsilon^*(g)(x) = \lim_{n \rinf}\varepsilon^*(g_{n})(x_{n}).$$

\noindent{\em Proof of Claim 1:}
There will be two cases 
\paragraph*{Case 1}$x \not \in \varepsilon[Y]$: Then $f(x) \neq 0$ and $f(x_{n}) \neq 0$ for almost all $n$. Then, locally around $x$, everything is continuous.
\paragraph*{Case 2} $x \in \varepsilon[Y]$: Then $\varepsilon^*(g)(x) = g(\varepsilon^{-1}(x))$. We may, without serious loss of generality, assume that for every $n \in \N$ we have that $x_{n} \not \in \varepsilon[Y]$ (since $g$ is continuous on $Y$ and $g = \lim_{n \rinf}g_{n}$ as functions defined on $Y$ in the limit space sense).
Then  
\[\varepsilon^*(g_{n})(x_{n}) =(1-\lambda_{n})\!\!\!\!\sum_{a \in A_{m_{n}}}\!\!\mu_{m_{n}}(x_{n})(a)\cdot g_{n}(\nu_{m_{n}}(a)) + \lambda_{n}\!\!\!\!\!\sum_{b \in A_{m_{n}+1}}\!\!\mu_{m_{n}+1}(x_{n})(b)\cdot g_{n}(\nu_{m_{n}+1}(b))\] where $m_{n} \in \N$ and $\lambda_{n} \in [0,1)$ are such that $f(x_{n}) = \frac{1}{m_{n} + \lambda_{n}}$.

Now, if we for each $n$ select $a_{n}$ such that $a_{n} \in A_{m_{n}}$ and $\mu_{m_{n}}(x_{n})(a_{n}) > 0$ or such that $a_{n} \in A_{m_{n}+1}$ and $\mu_{m_{n}+1}(x_{n})(a_{n}) > 0$, we may use that $x = \lim_{n \rinf} x_{n}$ and the properties of probabilistic projections to see that $\varepsilon^{-1}(x) = \lim_{n \rinf}\nu_{m_n/m_{n}+1}(a_{n})$, where we choose the index $m_{n}$ or $m_{n}+1$ that is relevant for $a_{n}$.

Then $g(\varepsilon^{-1}(x)) = \lim_{n \rinf}g_{n}(\nu_{m_{n}/m_{n}+1}(a_{n}))$ for each such sequence.

Since $\varepsilon^*(g_{n})(x_{n})$ is a weighted sum of values $g_{n}(\nu_{m_{n}}(a))$ for $a \in A_{m_{n}}$ and $g_{n}(\nu_{m_{n} + 1}(a))$ for $a \in A_{m_{n}+1}$, where the sum of the coefficients is 1 and the coefficients are given by the probabilities derived from $x_{n}$, it follows from the consideration above that $$\varepsilon^*(g)(x) = g(\varepsilon^{-1}(x)) = \lim_{n \rinf}\varepsilon^*(g_{n})(x_{n}).$$
This ends the proof of Claim 1. 

Note that $(\varepsilon^*)^{-1}(\gamma)$ defined by
$$(\varepsilon^*)^{-1}(\gamma)(y) = \gamma(\varepsilon(y))$$ will map $X \rightarrow U$ onto $Y \rightarrow U$, and that $(\varepsilon^*)^{-1}$ will be the inverse of $\varepsilon^*$ on the image of $\varepsilon^*$. Thus $\varepsilon^*$ is a homeomorphism onto its range.\medskip

\noindent{\em Claim 2:}
There is a continuous $$h:(X \rightarrow U) \rightarrow [0,1]$$ such that $h^{-1}(\{0\})$ is the range of $\varepsilon^*$.\medskip

\noindent{\em Proof of Claim 2:}
Let $\{y_{n}\}_{n \in \N}$ be a dense subset of $Y$ and $\{x_{m}\}_{m \in \N}$ a dense subset of $X$.

Given $\gamma:X \rightarrow U$ we will let $h(\gamma)$ measure to what extent $\gamma$ does not map $\varepsilon[Y]$ into $A$ and to what extent $\gamma$ will differ from $\varepsilon^*((\varepsilon^*)^{-1}(\gamma))$.

Note that the definition of $(\varepsilon^*)^{-1}(\gamma)$ makes sense since we never use that a function  takes values in $A$ in the definition or in the proof of Claim 1.

We simply let
$$h(\gamma) = \sum_{n \in \N}2^{-(n+1)}(min\{1,d_{U}(A, \gamma(\varepsilon(y_{n})) )+ d_{U}(\gamma(x_{n}),\varepsilon^*((\varepsilon^*)^{-1}(\gamma))(x_{n}))\}).$$
This ends the proof of Claim 2.

It remains to produce the probabilistic projection. 
Let $\Ps$ be a countable pseudobase for $Y$, see Section \ref{section4}. Let $\{\xi_{n}\mid n \in \N\}$ be a countable dense subset of $U$. For $r > 0$, $r \in \Q$, we let 
$$B_{n,r} = \{a \in U \mid d_{A}(a,\xi_{n}) \leq r\}.$$

Let $\{\langle p_{i},B_{i}\rangle\}_{i \in \N}$ be an enumeration of all pairs $\langle p,B\rangle$ where $p \in \Ps$ and $B$ is a nonempty finite intersection of closed neighborhoods of the form $B_{n,r}$.

We say that $\langle p_{i},B_{i}\rangle$ {\em approximates} $\gamma \in X \rightarrow U$ if $\gamma(y) \in B_{i}$ whenever $y \in p_{i}$, cf. the construction of pseudobase elements for function spaces.

Let $K \subseteq \N$ be finite. $K$ is {\em relevant} if there is a $g:Y \rightarrow A$ such that $$ \ast\;\;\forall i \in K \forall y \in p_{i }(g(y) \in B_{i}).$$ 
If $K$ is relevant, let $g_{K}$ satisfy $\ast$.

If $K$ is not relevant, let $m$ be maximal such that $K \cap \{1 , \ldots , m\}$ is relevant, and let $$g_{K} = g_{K \cap \{1 , \ldots m\}}.$$
Now, we assume that the enumeration $\{y_{j}\}_{j \in \N}$ of the dense subset of $Y$ used in the proof of Claim 2 is chosen such that for all $p \in \Ps$, $\{y_{j}\mid y_{j} \in p\}$ is a dense subset of $p$. Then, whenever $p \in \Ps$, $B\subseteq A$ is a closed set and $g:Y \rightarrow A$ is continuous we have that
$$\forall y \in p(g(y) \in B) \Leftrightarrow \forall j \in \N(y_{j} \in p \Rightarrow g(y_{j}) \in B).$$
Now, let $C_{n}$ be the powerset of $\{1 , \ldots , n\}$.
We will construct a sequence of continuous functions $$\mu^*_{n}:(X \rightarrow U) \rightarrow PD(C_{n}).$$ 
Let $k = k_{n}$ be so large that for all $i \leq n$ there is a $j \leq k$ such that $y_{j} \in p_{i}$.
\begin{enumerate}[$-$]
\item
Let $\mu^*_{n,i}(\gamma)(\in) = 1$ if $\gamma(\varepsilon(y_{j})) \in B_{i}$ for all $j \leq k$  with $y_{j} \in p_{i}$.
\item
Let $\mu^*_{n,i}(\gamma)(\in) = 0$ if $d_{U}(B_{i} ,\gamma(\varepsilon(y_{j}))) \geq 2^{-n}$ for at least one $j \leq k$ with $y_{j} \in p_{i}$
\item
Let $\mu^*_{n,i}(\gamma)(\in) = 1-\lambda$ if $$2^{-n}\cdot \lambda = max\{d_{U}(B_{i}, \gamma(\varepsilon(y_{j})))\mid j \leq k \wedge y_{j} \in p_{i}\}$$ otherwise.
\item Let $\mu^*_{n,i}(\gamma)(\notin) = 1 - \mu^*_{n,i}(\gamma)(\in)$.
\end{enumerate}
$\mu_{n,i}^*(\gamma)$ is a probability distribution on the two-point set $\{\in,\notin\}$ where the probability of $\in$ is measuring how probable it is, given $n$, that $\langle p_{i},B_{i}\rangle$ approximates $\gamma$.

Let $$\mu^*_{n}(\gamma)(K) = \prod_{i \in K}\mu^*_{n,i}(\gamma)(\in) \cdot \prod_{i \not \in K}\mu^*_{n,i}(\gamma)(\not \in).$$
This gives us the $n$'th estimate of how likely it is that $K$ is the set of indices of the approximations to $\gamma$.\medskip

\noindent{\em Claim 3:}
Assume that $g:Y \rightarrow A$, $\gamma = \varepsilon^*(g)$ and that $\gamma = \lim_{n \rinf}\gamma_{n}$.
Assume further that for each $n \in \N$, $K_{n} \in C_{n}$ is such that $\mu^*_{n}(\gamma_{n})(K_{n}) > 0$.
Then $g = \lim_{n \rinf}g_{K_{n}}$.\medskip

\noindent{\em Proof of Claim 3:}
Using the lim-space characterization it is sufficient to show that whenever $z = \lim_{n \rinf}z_{n } \in Y$, then $g(z) = \lim_{n \rinf}g_{K_{n}}(z_{n})$ in $U$.
We will use Lemma \ref{lemma1}.

Let $(D, D^R, \delta)$ be the admissible domain representation of $Y$, where $D$ consists of ideals of pseudobase elements in $\Ps$, and let $(E,E^R,\delta_{1})$ be the corresponding domain representation of $Y \rightarrow U$, see the proof of Lemma \ref{lemma22} for the construction and the notation used below.

Let $\alpha = \lim_{n \rinf} \alpha_{n}$ be a convergent sequence from $E^R$ representing $$g = \lim_{n \rinf}(\varepsilon^*)^{-1}(\gamma_{n})$$ and let $\zeta = \lim_{n \rinf} \zeta_{n}$ be a convergent sequence from $D^R$ representing $z = \lim_{n \rinf}z_{n}$, see Remark \ref{remark7}.

Let $\epsilon > 0$. Since $\alpha$ represents $g$ and $\zeta$ represents $z$, there is an $m \in \N$ such that $P_{\{\langle p_{m},B_{m}\rangle\}} \in \alpha$, $p_{m} \in \zeta$ and such that the diameter of $B_{m}$ is less than $\epsilon$. We will show that for sufficiently large $n$ we have that $g_{K_{n}}(z_{n}) \in B_{m}$. This will show the claim.

Let $n_{0}$ be such that for $n \leq n_{0}$ we have that $P_{\{\langle p_{m},B_{m}\rangle\}} \in \alpha_{n}$ and that $p_{m} \in \zeta_{n}$.

Recall how we used $k_{n}$ in the construction of $\mu^*_{n}(g)$. Let $n_{1}$ be so large that for any $i \leq m$, if $g[p_{i}] \not \subseteq B_{i}$, then there is a $j \leq k_{n_{1}}$ such that $y_{j} \in p_{i}$ and $g(y_{j} ) \not \in B_{i}$.

Select one such $j_{i}$ for each relevant $i \leq m$, and then choose $n_{2}$ so large that for each $n \geq n_{2}$ and each relevant $i \leq m$ we have that
$$d_{U}(\gamma_{n}(\varepsilon(y_{j_{i}})),B_{i}) > 2^{-n}.$$
This is possible since $\gamma(\varepsilon(y_{j_{i}})) = \lim_{n \rinf }\gamma_{n}(\varepsilon(y_{j_{i}}))$.

Let $n \geq max\{n_{0},n_{1},n_{2}\}$ and let $K \subseteq \{1, \ldots , n\}$ be such that $\mu^*_{n}(\gamma_{n})(K) > 0$.

For $i < m$ we have ensured that if $\gamma_{n}[\varepsilon[p_{i}]] \not \subseteq B_{i}$, then $\mu^*_{n,i}(\gamma_{n})(\in) = 0$ and since $P_{\{\langle p_{m},B_{m}\rangle} \} \in \alpha_{n}$ we also have that $\mu^*_{n,m}(\gamma_{n})(\in) = 1$.
It follows that $g$ witnesses that $K \cap \{1, \ldots , m\}$ is relevant and contains $m$. This holds in particular for $K = K_n$, so $g_{K_{n}}(z_{n}) \in B_{m}$. This ends the proof of Claim 3.

Now the proof of Lemma \ref{lemma6.2} is complete, but let us summarize what we have achieved.
\begin{enumerate}[$-$]
\item We have defined the embedding $\varepsilon^*:(Y \rightarrow A) \rightarrow (X \rightarrow U)$ and proved that it is continuous and has a continuous inverse on its range.
\item We have proved that the range of $\varepsilon^*$ is a functionally closed set.
\item We have defined the finite set $C_{n}$ and the map $$K \mapsto g_{K}$$ from $C_{n}$ into $Y \rightarrow A$. Let $\nu^*_{n}(K) = g_{K}$.
\item For each $\gamma \in X \rightarrow U$, we have defined the probability distribution $\mu^*_{n}(\gamma)$ on $C_{n}$ and proved that altogether, $\varepsilon^*$ and $\{\langle C_{n},\nu^*_{n},\mu^*_{n}\rangle\}_{n \in \N}$ form  a probabilistic embedding-projection pair between $Y \rightarrow A$ and $X \rightarrow U$.\qed
\end{enumerate}

\noindent We have not included cartesian products as one type
constructor, but in order to handle types of the form $\sigma = \tau
\rightarrow \delta$ in the reflection of Lemma \ref{lemma6.2} it will
make life simpler if we view any type $\sigma$ as a type $\sigma =
\tau_{1} , \ldots , \tau_{m} \rightarrow V_{i}$ where $V_{i}$ is
interpreted as some separable metric space. This means that we need an
extra induction step in the proof of Theorem \ref{theorem6.1}, the
case of products.

If $X_{1} , \ldots , X_{m}$ are spaces in $\mathcal Q$ or in $qcb$ in general, the product $\prod_{i = 1}^m X_{i}$ is not just the standard topological product, but the finest topology accepting the induced convergent sequences in the product topology as convergent.  We then have
\begin{lemma} \label{lemma6.3} Let $Y_{1} , \ldots , Y_{m}$ and $X_{1} , \ldots , X_{m}$ be two sequences of spaces in $\mathcal Q$, and assume that there are probabilistic embedding-projection pairs between $Y_{i}$ and $X_{i}$ for each $i \leq m$.
Then there is a probabilistic embedding-projection pair between $\prod_{i = 1}^mY_{i}$ and $\prod_{i = 1}^mX_{i}$.
\end{lemma}
\proof
This is more an observation than a lemma:
\begin{enumerate}[$-$]
\item If $\varepsilon_{i}$ is the embedding for each $i \leq m$,  we let $\varepsilon = \prod_{i = 1}^m\varepsilon_{i}$.
\item If $f_{i}$ witnesses that the range of $\varepsilon_{i}$ is a functionally closed set for each $i \leq m$, let $$f(x_{1}, \ldots , x_{m} )= \frac{1}{m}\sum_{i = 1}^m f_{i}(x_{i})$$ witness that the range of $\varepsilon$ is a functionally closed set.
\item If $A_{n}^k$ and $\nu_{n}^k:A_{n}^k \rightarrow Y_{i}$ are the finite ``approximations'' to $Y_{i}$ used for the probabilistic projections, we let $A_{n}$ and $\nu_{n}$ be obtained by just taking products.
\item The probability distributions of the product are just the products of the probability distributions of each coordinate.
\end{enumerate}
It is easy to verify that all properties are preserved in this construction.
\qed
Now we have all the ingredients needed to prove Theorem \ref{theorem6.1}:

If $X_{1}, \ldots , X_{k}$ are complete, separable metric spaces, and $\sigma$ is a type expression in the variables $V_{1}, \ldots , V_{k}$, we prove by induction on $\sigma$ that there is a probabilistic embedding-projection pair between $\sigma(X_{1} , \ldots , X_{k})$ and $\sigma(U , \ldots , U)$, where the image of the embedding is a functionally closed set.

The induction start $\sigma = V_{i}$ is covered by Lemma \ref{lemma6.2new}.

For the induction step, we let $\sigma = \tau_{1} , \ldots , \tau_{m} \rightarrow V_{j}$.

We then use Lemma \ref{lemma6.3} and the induction hypothesis to show that there is a probabilistic embedding-projection pair between $$\prod_{i = 1}^m\tau_{i}(X_{1} , \ldots, X_{k})\qquad\hbox{and}\qquad\prod_{i = 1}^m \tau_{i}(U,\ldots,U).$$
We then use Lemma \ref{lemma6.2} to complete the induction step.
\qed
\begin{remark}{\em This proof is noneffective. We have used that $U$ is homeomorphic to $l_{2}$, and we do not know of any effective proof of that. There are likely to be methods that  get us around this problem, using effective semiconvexity like we did in Section \ref{section5}.

However, the concept of a relevant set of natural numbers, and the choice of the functions $g_{K}$ in the proof of Lemma \ref{lemma6.2}, are not effective in a general situation, even when the metric spaces $X_{1} , \ldots , X_{k}$ are effective. Thus we may as well use the topological characterization of $U$ as homeomorphic to $l_{2}$ in this proof.}\end{remark}
\begin{remark}{\em If we let $\varepsilon_{V_i}$ be the isometric map from $X_i$ to $U$ used in this proof, we in reality construct, in the proof of Theorem \ref{theorem6.1}, an embedding $$\varepsilon_\sigma:\sigma(X_1, \ldots , X_k) \rightarrow \sigma(U,\ldots,U)$$ by recursion on $\sigma$. Actually, we construct an embedding of the typed hierarchy over $X_1, \ldots,X_k$ to the corresponding hierarchy over $U,\ldots,U$ in the sense that our local embeddings commute with application in the two hierarchies. We did not stress this in the proof, and leave it as an observation.}\end{remark}

\section{Conclusions and further research}\label{section8}
\noindent We have shown that the typed hierarchy of hereditarily continuous and total functionals over the Urysohn space $U$ is rich enough to contain all typed hierarchies over separable metric spaces as topological sub-hierarchies. One problem is if this can be generalized to a situation where we do not consider only the full space of continuous functions at types $\sigma = \tau \rightarrow \delta$, but also cases where we select a functionally closed subset of the set of all continuous functions. If we work within the category of $qcb$-spaces with a pseudobase of functionally closed sets, see Schr\"{o}der \cite{Math}, we may apply his result stating that functionally closed in functionally closed is functionally closed, and our embedding theorem should also be valid in this generalized context. We consider this as a conjecture since we have not worked out a detailed proof.

All our spaces $\sigma(\vec X)$ are homeomorphic to functionally closed subsets of spaces of the form $X \rightarrow U$ where $X \in {\mathcal Q}$, but we have not studied this class, denoted by $zero({\mathcal Q} \rightarrow U)$, more closely. (Recall that these sets are also called {\em zero-sets}.) P. K.  K\o ber \cite{Petter} has obtained some partial results related to strictly positive inductive definitions of topological spaces, and one consequence of his results is that there is a least fixed point of a strictly positive inductive definition with parameters from $zero({\mathcal Q} \rightarrow U)$ in $
zero({\mathcal Q} \rightarrow U)$ itself.

We know that $U$ is homeomorphic to $l_{2}$, but we do not know if the $l_{2}$-structure on $U$ is effective in the sense that there are computable
\[|| \;\; ||:U \rightarrow \R_{\geq 0}\qquad
+:U \times U \rightarrow U\qquad
\cdot : \R \times U \rightarrow U
\]
representing the $l_{2}$ - structure, or any other Banach space structure on $U$.

It may be of interest to equip $U$ with some structure offering an internal computability theory, e.g.\  by identifying subsets representing $\N$, $\Z$ and $\R$.

\section*{Acknowledgments}
\noindent I am grateful for discussions with (in alphabetic order) Philipp Gerhardy, Petter K\o ber, Yiannis Moschovakis, Mathias Schr\"oder and Vladimir Uspenskij on topics with relevance to this paper.

I  am also grateful for the constructive remarks made by the three referees.

\end{document}